\documentclass[aps,preprint]{revtex4}
\usepackage[dvips]{graphicx}
\begin{document}
\draft
\title{\bf Experimental investigation of nodal domains in the chaotic microwave rough billiard}
\author{Nazar Savytskyy, Oleh Hul and Leszek Sirko}

\address{Institute of Physics, Polish Academy of Sciences, Aleja Lotnik\'{o}w 32/46, 02-668 Warszawa, Poland}
\date{August 26, 2004}

\bigskip

\begin{abstract}
We present the results of experimental study of nodal domains of
wave functions (electric field distributions) lying in the regime
of Shnirelman ergodicity in the chaotic half-circular microwave
rough billiard. Nodal domains are regions where a wave function
has a definite sign. The wave functions $\Psi_N$  of the rough
billiard were measured up to the level number $N=435$. In this way
the dependence of the number of nodal domains $\aleph_N$ on the
level number $N$ was found. We show that in the limit $N
\rightarrow \infty $ a least squares fit of the experimental data
reveals the asymptotic number of nodal domains $\aleph_N/N \simeq
0.058 \pm 0.006$ that is close to the theoretical prediction
$\aleph_N/N \simeq 0.062$. We also found that the distributions of
the areas $s$ of nodal domains and their perimeters $l$  have
power behaviors $n_s \propto s^{-\tau}$ and $n_l \propto
l^{-\tau'}$, where scaling exponents are equal to $\tau = 1.99 \pm
0.14$ and $\tau'=2.13 \pm 0.23$, respectively. These results are
in a good agreement with the predictions of percolation theory.
Finally, we demonstrate that for higher level numbers $N \simeq
220-435$ the signed area distribution oscillates around the
theoretical limit $\Sigma_{A} \simeq 0.0386N^{-1}$.
\end{abstract}


\bigskip
\maketitle

\smallskip

In recent papers Blum {\it et al.} \cite{Blum2002} and Bogomolny
and Schmit \cite{Bogomolny2002} have considered the distribution
of the nodal domains of real wave functions $\Psi(x,y)$ in 2D
quantum systems (billiards). The condition $\Psi(x,y)=0$
determines a set of nodal lines which separate regions (nodal
domains) where a wave function $\Psi(x,y)$ has opposite signs.
Blum {\it et al.} \cite{Blum2002} have shown that the
distributions of the number of nodal domains can be used to
distinguish between systems with integrable and chaotic underlying
classical dynamics. In this way they provided a new criterion of
quantum chaos, which is not directly related to spectral
statistics. Bogomolny and Schmit \cite{Bogomolny2002} have shown
that the distribution of nodal domains for quantum wave functions
of chaotic systems is universal. In order to prove it they have
proposed a very fruitful, percolationlike, model for description
of properties of the nodal domains of generic chaotic system. In
particular, the model predicts that the distribution of the areas
$s$ of nodal domains should have power behavior $n_s \propto
s^{-\tau}$, where $\tau=187/91$ \cite{Ziff1986}.

In this paper we present the first experimental investigation of
nodal domains of wave functions of the chaotic microwave rough
billiard. We tested experimentally some of important findings of
papers by Blum {\it et al.} \cite{Blum2002} and Bogomolny and
Schmit \cite{Bogomolny2002} such as the signed area distribution
$\Sigma_{A}$ or the dependence of the number of nodal domains
$\aleph_N$ on the level number $N$. Additionally, we checked the
power dependence of nodal domain perimeters $l$, $n_l \propto
l^{-\tau'}$, where according to percolation theory the scaling
exponent $\tau'=15/7$ \cite{Ziff1986}, which was not considered in
the above papers.

\begin{figure}[!]
\begin{center}
\rotatebox{270} {\includegraphics[width=0.5\textwidth,
height=0.6\textheight, keepaspectratio]{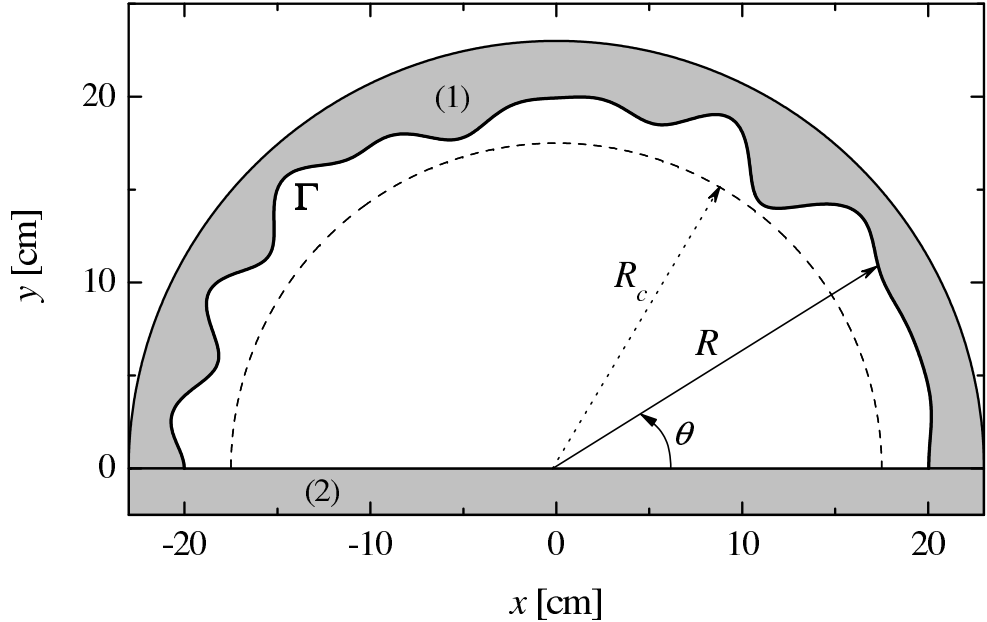}} \caption{Sketch of
the chaotic half-circular microwave rough billiard in the $xy$
plane. Dimensions are given in cm. The cavity sidewalls are marked
by 1 and 2 (see text). Squared wave functions $|\Psi_N(R_c,\theta
)|^2$  were evaluated on a half-circle of  fixed radius $R_c=17.5$
cm. Billiard's rough boundary $\Gamma $ is marked with the bold
line.}\label{Fig1}
\end{center}
\end{figure}

In the experiment we used the thin (height $h=8$ mm) aluminium
cavity in the shape of a rough half-circle (Fig. \ref{Fig1}). The
microwave cavity simulates the rough quantum billiard due to the
equivalence between the Schr\"odinger equation and the Helmholtz
equation \cite{Hans,Hans2}. This equivalence remains valid for
frequencies less than the cut-off frequency  $\nu_c =c/2h \simeq
18.7$ GHz, where c is the speed of light. The cavity sidewalls are
made of 2 segments. The rough segment 1 is described by the radius
function
$R(\theta)=R_{0}+\sum_{m=2}^{M}{a_{m}\sin(m\theta+\phi_{m})}$,
where  the mean radius $R_0$=20.0 cm, $M=20$, $a_{m}$ and
$\phi_{m}$ are uniformly distributed on [0.269,0.297] cm and
[0,2$\pi$], respectively, and $0\leq\theta<{\pi}$.  It is worth
noting that following our earlier experience
\cite{Hlushchuk01b,Hlushchuk01} we decided to use a rough
half-circular cavity instead of a rough circular cavity because in
this way we avoided nearly degenerate low-level eigenvalues, which
could not be possible distinguished in the measurements. As we
will see below, a half-circular geometry of the cavity was also
very suitable in the accurate measurements of the electric field
distributions inside the billiard.

The surface roughness of a billiard is characterized by the
function $k(\theta)=(dR/d\theta)/R_0$. Thus for our billiard we
have the angle average $\tilde k=(\left <k^{2}(\theta)\right
>_{\theta})^{1/2}\simeq  0.488$. In such a billiard the dynamics is
diffusive in orbital momentum due to collisions with the rough
boundary because $\tilde k $ is much above the chaos border
$k_c=M^{-5/2}=0.00056$ \cite{Frahm97}. The roughness parameter
$\tilde k $ determines also other properties of the billiard
\cite{Frahm}. The eigenstates are localized for the level number
$N < N_e = 1/128 \tilde k^4$. Because of a large value of the
roughness parameter  $\tilde k $ the localization border lies very
low, $N_e \simeq 1$.  The border of Breit-Wigner regime is $N_W =
M^2/48\tilde k^2 \simeq 35$. It means that between $N_e < N < N_W$
Wigner ergodicity \cite{Frahm} ought to be observed and for $N >
N_W$ Shnirelman ergodicity should emerge.  In 1974 Shnirelman
\cite{Shnirelman} proved that quantum states in chaotic billiards
become ergodic for sufficiently high level numbers. This means
that for high level numbers wave functions have to be uniformly
spread out in the billiards. Frahm and Shepelyansky \cite{Frahm}
showed that in the rough billiards the transition from the
exponentially localized states to the ergodic ones is more
complicated and can pass through an intermediate regime of Wigner
ergodicity. In this regime the wave functions are nonergodic and
compose of rare strong peaks distributed over the whole energy
surface. In the regime of Shnirelman ergodicity the wave functions
should be distributed homogeneously on the energy surface.

In this paper we focus our attention on Shnirelman ergodicity
regime.

One should mention that rough billiards and related systems are of
considerable interest elsewhere, e.g.  in the context of dynamic
localization \cite{Sirko00}, localization in discontinuous quantum
systems \cite{Borgonovi}, microdisc lasers \cite{Yamamoto,Stone}
and ballistic electron transport in microstructures
\cite{Blanter}.

In order to investigate properties of nodal domains knowledge of
wave functions (electric field distributions inside the microwave
billiard) is indispensable. To measure the wave functions we used
a new, very effective  method described in \cite{Savytskyy2003}.
It is based on the perturbation technique and preparation of the
``trial functions". Below we will describe shortly this method.

 The  wave functions
$\Psi_N(r,\theta )$ (electric field distribution $E_N(r,\theta )$
inside the cavity) can be determined from the form  of electric
field $E_N(R_c, \theta )$ evaluated  on a half-circle of  fixed
radius $R_c$ (see Fig. \ref{Fig1}). The first step in evaluation
of $E_N(R_c, \theta )$ is measurement of $|E_N(R_c, \theta )|^2$.
The perturbation technique developed in \cite{Slater52} and used
successfully in \cite{Slater52,Sridhar91,Richter00,Anlage98} was
implemented for this purpose. In this method a small perturber is
introduced inside the cavity to alter its resonant frequency
according to $$ \nu -\nu_N =\nu_N(aB_N^2-bE_N^2), \eqno(1)$$ where
$\nu_N $ is the $N$th resonant frequency of the unperturbed
cavity, $a$ and $b$ are geometrical factors.  Equation (1) shows
that the formula can not be used to evaluate $E_N^2$ until the
term containing magnetic field $B_N$ vanishes. To minimize the
influence of $B_N$ on the frequency shift $\nu -\nu_N $ a small
piece of a metallic pin  (3.0 mm in length and 0.25 mm in
diameter) was used as a perturber. The perturber was moved by the
stepper motor via the Kevlar line hidden in the groove (0.4 mm
wide, 1.0 mm deep) made in the cavity's bottom wall along the
half-circle $R_c$. Using such a perturber we had no positive
frequency shifts that would exceed the uncertainty of frequency
shift measurements (15 kHz). We checked that the presence of the
narrow groove in the bottom wall of the cavity caused only very
small changes $\delta \nu_N$ of the eigenfrequencies $\nu_N$ of
the cavity $|\delta \nu_N|/\nu_N \leq 10^{-4}$. Therefore, its
influence into the structure of the cavity's wave functions was
also negligible. A big advantage of using  hidden in the groove
line was connected with the fact that the attached to the line
perturber was always vertically positioned what is crucial in the
measurements of the square of electric field $E_N$. To eliminate
the variation of resonant frequency connected with the thermal
expansion of the aluminium cavity  the temperature of the cavity
was stabilized with the accuracy of 0.05 $\deg $.

\begin{figure}[!]
\begin{center}
\rotatebox{0} {\includegraphics[width=0.5\textwidth,
height=0.8\textheight, keepaspectratio]{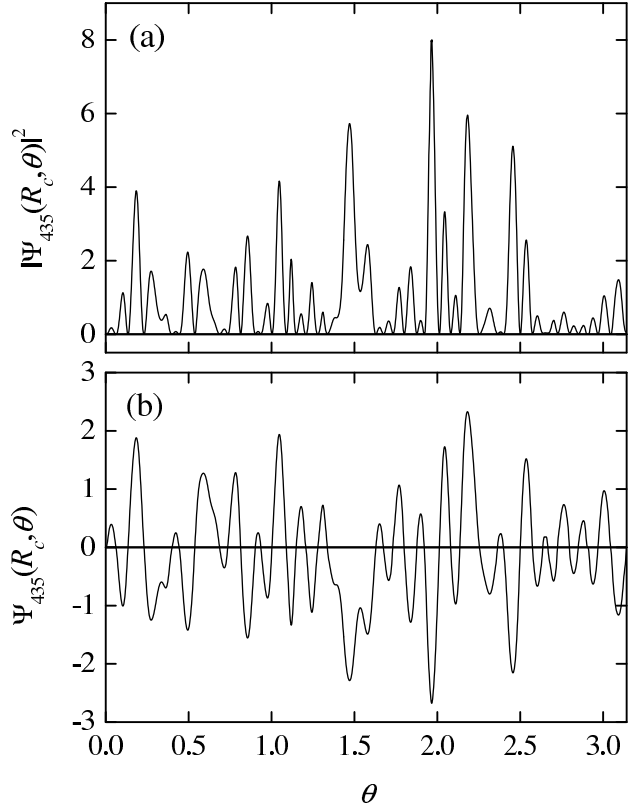}} \caption{Panel (a):
Squared wave function $|\Psi_{435}(R_c,\theta )|^2$ (in arbitrary
units) measured on a half-circle with radius $R_c=17.5$ cm
($\nu_{435} \simeq 14.44$ GHz). Panel (b): The ``trial wave
function" $\Psi_{435}(R_c,\theta )$ (in arbitrary units) with the
correctly assigned  signs, which was used in the reconstruction of
the wave function $\Psi_{435}(r, \theta )$ of the billiard (see
Fig. \ref{Fig3}). }\label{Fig2}
\end{center}
\end{figure}

The regime of  Shnirelman ergodicity for the experimental rough
billiard is defined for $N > 35$. Using a field perturbation
technique we measured squared wave functions $|\Psi_N(R_c,\theta
)|^2$  for 156 modes within the region $80\leq N \leq 435$. The
range of corresponding eigenfrequencies was from $\nu_{80} \simeq
6.44$ GHz to $\nu_{435} \simeq 14.44$ GHz.
 The measurements were performed at 0.36 mm steps along a half-circle with fixed
 radius $R_c=17.5$ cm. This step was small enough to reveal in details the space structure of high-lying levels.
In Fig. \ref{Fig2} (a) we show the example of the squared wave
function $|\Psi_N (R_c, \theta )|^2$ evaluated for the level
number $N=435$. The perturbation method used in our measurements
allows us to extract information about the wave function amplitude
$|\Psi_N(R_c, \theta )|$ at any given point of the cavity but it
doesn't allow to determine the sign of $\Psi_N(R_c, \theta )$
\cite{Stein95}. Our results presented in \cite{Savytskyy2003}
suggest the following sign-assignment strategy: We begin with the
identification of  all close to zero minima of  $|\Psi_N(R_c,
\theta )|$.  Then the sign ``minus" maybe arbitrarily assigned to
the region between the first and the second minimum, ``plus" to
the region between the second minimum and the third one, the next
``minus" to the next region between consecutive minima and so on.
In this way we construct our ``trial wave function" $\Psi_N(R_c,
\theta )$. If the assignment of the signs is correct we should
reconstruct the wave function $\Psi_N(r, \theta )$ inside the
billiard with the boundary condition $\Psi_N(r_{\Gamma },
\theta_{\Gamma } )=0$.

The  wave functions of a rough half-circular billiard may be
expanded in terms of circular waves (here only odd states in
expansion are considered)
$$
\Psi_N(r, \theta ) = \sum_{s=1}^L a_s C_s J_{s}(k_Nr)\sin(s\theta
), \eqno(2)
$$
where $C_s=(\frac{\pi
}{2}\int_0^{r_{max}}|J_{s}(k_Nr)|^2rdr)^{-1/2}$ and $k_N=2\pi
\nu_N /c$.

\begin{figure}[!]
\begin{center}
\rotatebox{270} {\includegraphics[width=0.5\textwidth,
height=0.6\textheight, keepaspectratio]{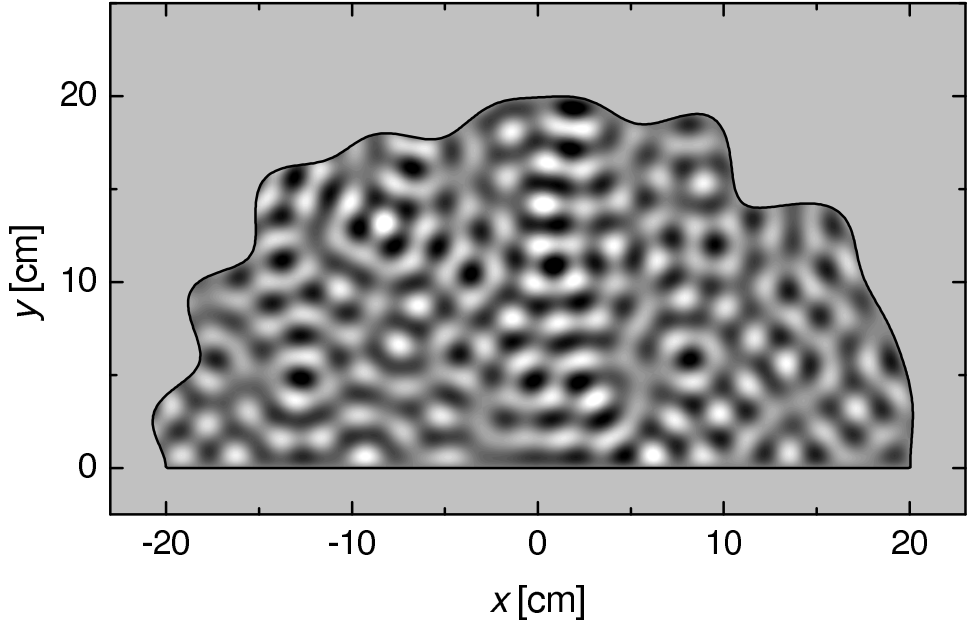}} \caption{ The
reconstructed wave function $\Psi_{435}(r,\theta ) $ of the
chaotic half-circular microwave rough billiard. The amplitudes
have been converted into a grey scale with white corresponding to
large positive and black corresponding to large negative values,
respectively. Dimensions of the billiard are given in cm.
}\label{Fig3}
\end{center}
\end{figure}

In Eq. (2) the number of basis functions is limited to $L=k_N
r_{max} = l_{N}^{max}$, where $r_{max}=21.4$ cm is the maximum
radius of the cavity. $l_N^{max} = k_N r_{max}$ is a semiclassical
estimate for the maximum possible angular momentum for a given
$k_N$. Circular waves with angular momentum $s > L$ correspond to
evanescent waves and can be neglected. Coefficients $a_s$ may be
extracted from the ``trial wave function" $\Psi_N(R_c, \theta )$
via
$$
a_s=[\frac{\pi }{2}C_sJ_{s}(k_NR_c)]^{-1}\int_0^{\pi}\Psi_N
(R_c,\theta )\sin(s\theta )d\theta . \eqno(3)
$$
Since our ``trial wave function" $\Psi_N (R_c,\theta )$ is only
defined on a half-circle of fixed radius $R_c$ and is not
normalized we imposed normalization of the coefficients $a_s$:
$\sum_{s=1}^{L}|a_s|^2 = 1$. Now, the  coefficients  $a_s$ and
Eq.~(2) can be used to reconstruct the wave function $\Psi_N(r,
\theta )$ of the billiard. Due to experimental uncertainties and
the finite step size in the measurements of $|\Psi_N(R_c, \theta
)|^2$ the wave functions $\Psi_N(r, \theta )$ are not exactly zero
at the boundary $\Gamma $.  As the quantitative measure of the
sign assignment quality we chose  the integral $\gamma
\int_{\Gamma }|\Psi_N(r,\theta )|^2dl $ calculated along the
billiard's rough boundary $\Gamma $, where $\gamma $ is length of
$\Gamma $. In Fig. \ref{Fig2} (b) we show the ``trial wave
function" $\Psi_{435} (R_c, \theta )$ with the correctly assigned
signs, which was used in the  reconstruction of the wave function
$\Psi_{435}(r, \theta )$ of the billiard (see Fig. \ref{Fig3}).
Using the method of the ``trial wave function" we were able to
reconstruct 138 experimental wave functions of the rough
half-circular billiard with the level number $N$ between 80 and
248 and 18 wave functions with $N$ between 250 and 435. The wave
functions were reconstructed on points of a square grid of side
$4.3 \cdot 10^{-4} $ m.  The remaining wave functions from the
range $N=80-435$ were not reconstructed because of the accidental
near-degeneration of the neighboring states or due to the problems
with the measurements of $|\Psi_N(R_c, \theta )|^2$ along a
half-circle coinciding for its significant part with one of the
nodal lines of $\Psi_N(r, \theta )$. These problems are getting
much more severe for $N>250$. Furthermore,  the computation time
$t_r$ required for reconstruction of the "trial wave function"
scales like $t_r \propto 2^{n_z -2}$, where $n_z$ is the number of
identified zeros in the measured function $|\Psi_N(R_c, \theta
)|$. For higher $N$, the computation time $t_r$  on a standard
personal computer with the processor AMD Athlon XP 1800+ often
exceeds several hours, what significantly slows down the
reconstruction procedure.

\begin{figure}[!]
\begin{center}
\rotatebox{0} {\includegraphics[width=0.5\textwidth,
height=0.8\textheight, keepaspectratio]{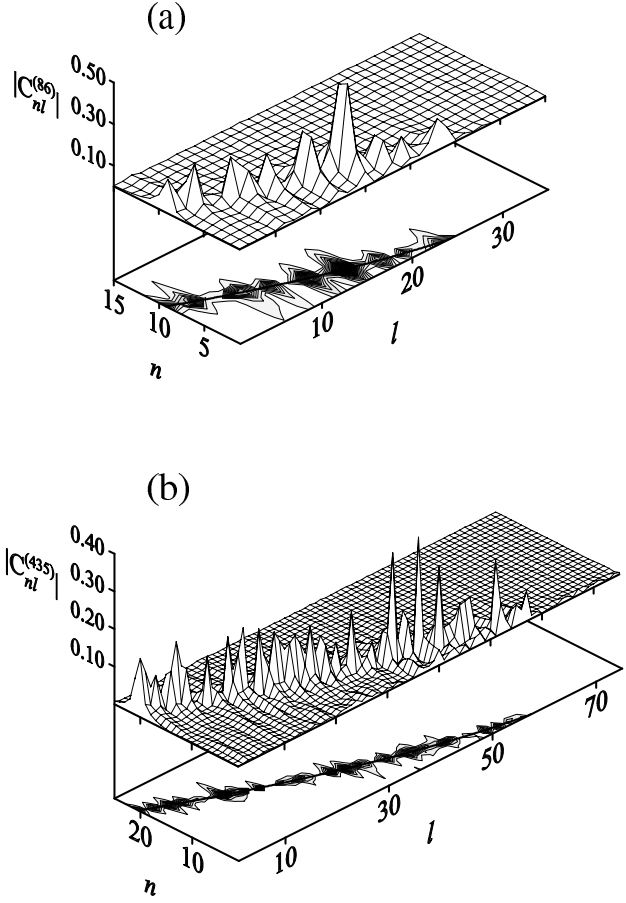}} \caption{Structure
of the energy surface in the regime of Shnirelman ergodicity.
 Here we show the moduli of amplitudes $|C^{(N)}_{nl}|$
for the wave functions: (a) $N=86$, (b) $N=435$. The wave
functions are delocalized in the $n, l$ basis. Full lines show the
semiclassical estimation of the energy surface (see
text).}\label{Fig4}
\end{center}
\end{figure}

Ergodicity of the billiard's wave functions can be checked by
finding the structure of the energy surface \cite{Frahm97}. For
this reason we extracted wave function amplitudes
$C^{(N)}_{nl}=\left< n,l|N \right>$ in the basis $n, l$ of a
half-circular billiard with radius $r_{max}$, where $ n=1,2,3
\ldots$ enumerates the zeros of the Bessel functions and $ l=1,2,3
\ldots$ is the angular quantum number. The moduli of amplitudes
$|C^{(N)}_{nl}|$  and their projections into the energy surface
for the representative experimental wave functions $N=86$ and
$N=435$ are shown in Fig. \ref{Fig4}. As expected, in the regime
of Shnirelman ergodicity the wave functions are extended
homogeneously over the whole energy surface \cite{Hlushchuk01}.
The full lines on the projection planes in Fig. \ref{Fig4}(a) and
Fig. \ref{Fig4}(b) mark the energy surface of a half-circular
billiard $H(n,l)=E_N=k^2_N$ estimated from the semiclassical
formula \cite{Hlushchuk01b}: $\sqrt{(l^{max}_{N})^2 - l^2}
-l\arctan(l^{-1}\sqrt{(l^{max}_{N})^2 - l^2}) + \pi/4 = \pi n $.
The peaks $|C^{(N)}_{nl}|$ are spread almost perfectly along the
lines marking the energy surface.

\begin{figure}[!]
\begin{center}
\rotatebox{0} {\includegraphics[width=0.5\textwidth,
height=0.8\textheight, keepaspectratio]{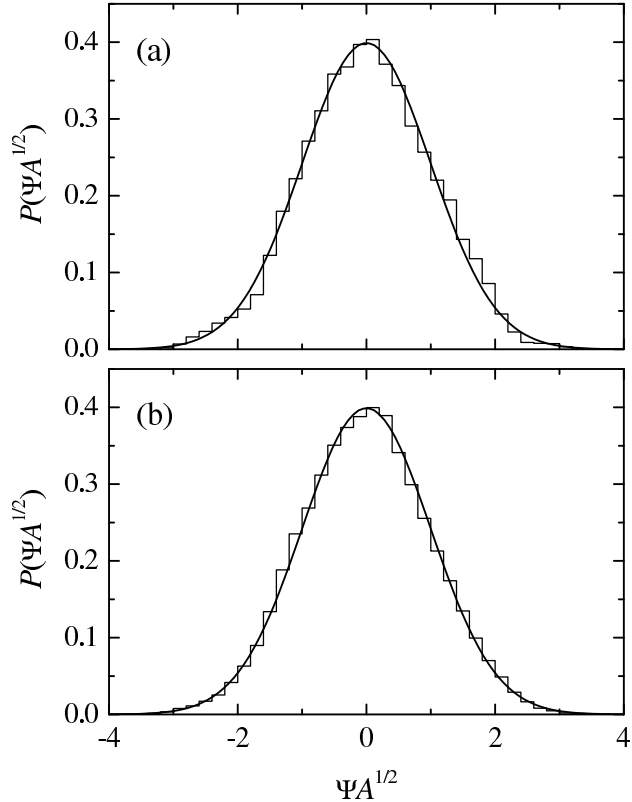}} \caption{Amplitude
distribution $P(\Psi A^{1/2})$ for the eigenstates: (a) $N=86$ and
(b) $N=435$ constructed as histograms with bin equal to 0.2. The
width of the distribution $P(\Psi)$ was rescaled to unity by
multiplying normalized to unity wave function by the factor
$A^{1/2}$, where $A$ denotes billiard's area. Full line shows
standard normalized Gaussian prediction $P_{0}(\Psi
A^{1/2})=(1/\sqrt{2\pi})e^{-\Psi^{2} A/2}$. }\label{Fig5}
\end{center}
\end{figure}

An additional confirmation of ergodic behavior of the measured
wave functions can be also sought in the form of the amplitude
distribution $P(\Psi)$ \cite{Berry77,Kaufman88}. For irregular,
chaotic states the probability of finding the value $\Psi$ at any
point inside the billiard, without knowledge of the surrounding
values, should be distributed as a Gaussian, $P(\Psi) \sim
e^{-\beta \Psi^{2}}$.   It is worth noting that in the above case
the spatial intensity should be distributed according to
Porter-Thomas statistics \cite{Hans2}.  The amplitude
distributions $P(\Psi A^{1/2})$ for the wave functions $N=86$ and
$N=435$ are shown in Fig. \ref{Fig5}. They were constructed as
normalized to unity histograms with the bin equal to 0.2. The
width of the amplitude distributions $P(\Psi )$ was rescaled to
unity by multiplying normalized to unity wave functions by the
factor $A^{1/2}$, where $A$ denotes billiard's area (see formula
(23) in \cite{Kaufman88}). For all measured wave functions in the
regime of Shnirelman ergodicity  there is a good agreement with
the standard normalized Gaussian prediction $P_{0}(\Psi
A^{1/2})=(1/\sqrt{2\pi})e^{-\Psi^{2} A/2}$.

\begin{figure}[!]
\begin{center}
\rotatebox{270} {\includegraphics[width=0.5\textwidth,
height=0.6\textheight, keepaspectratio]{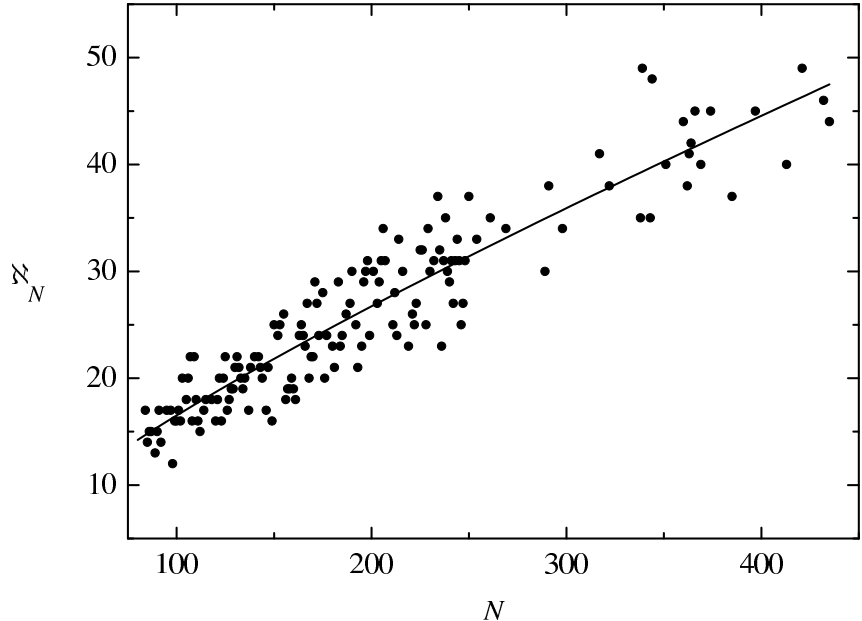}} \caption{The number
of nodal domains $\aleph_N$ (full circles) for the chaotic
half-circular microwave rough billiard. Full line shows a least
squares fit $\aleph_N = a_1N +b_1\sqrt{N}$ to the experimental
data (see text), where $a_1=0.058 \pm 0.006$, $b_1=1.075 \pm
0.088$. The prediction of the theory of Bogomolny and Schmit
\cite{Bogomolny2002} $a_1=0.062$. }\label{Fig6}
\end{center}
\end{figure}

The number of nodal domains $\aleph_N$ vs. the level number $N$ in
the chaotic microwave rough billiard is plotted in Fig.
\ref{Fig6}. The full line in Fig. \ref{Fig6} shows a least squares
fit $\aleph_N = a_1N +b_1\sqrt{N}$ of the experimental data, where
$a_1=0.058 \pm 0.006$, $b_1=1.075 \pm 0.088$. The coefficient
$a_1=0.058 \pm 0.006$ coincides with the prediction of the
percolation model of Bogomolny and Schmit \cite{Bogomolny2002}
$\aleph_N/N \simeq 0.062$ within the error limits. The second term
in a least squares fit corresponds to a contribution of boundary
domains,  i.e. domains, which include the billiard boundary.
Numerical calculations of Blum {\it et al.} \cite{Blum2002}
performed for the Sinai and stadium billiards showed that the
number of boundary domains scales as the number of the boundary
intersections, that is as $\sqrt{N}$. Our results clearly suggest
that in the rough billiard, at low level number $N$, the boundary
domains also significantly influence the scaling of the number of
nodal domains $\aleph_N$, leading to the departure from the
predicted scaling $\aleph_N \sim N$.

\begin{figure}[!]
\begin{center}
\rotatebox{270} {\includegraphics[width=0.5\textwidth,
height=0.6\textheight, keepaspectratio]{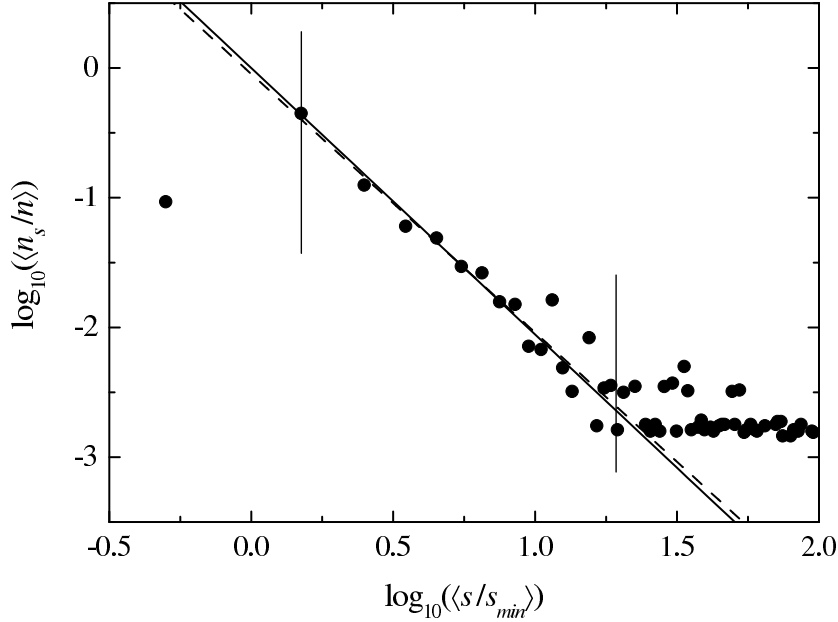}}
\caption{Distribution of nodal domain areas. Full line shows the
prediction of percolation theory $\log_{10}(\langle n_s/n \rangle)
= -\frac{187}{91} \log_{10}(\langle s/s_{min} \rangle)$. A least
squares fit $\log_{10}(\langle n_s/n \rangle) = a_2 -\tau
\log_{10}(\langle s/s_{min} \rangle)$ of the experimental results
lying within the vertical lines yields the scaling exponent $\tau
= 1.99 \pm 0.14$ and $a_2=-0.05 \pm 0.04$.  The result of the fit
is shown by the dashed line. }\label{Fig7}
\end{center}
\end{figure}

The bond percolation model \cite{Bogomolny2002} at the critical
point $p_c = 1/2$ allows us to apply other results of percolation
theory to the description of nodal domains of chaotic billiards.
In particular, percolation theory predicts that the distributions
of the areas $s$ and the perimeters $l$  of nodal clusters should
obey the scaling behaviors:  $n_s \propto s^{-\tau}$ and $n_l
\propto l^{-\tau'}$, respectively. The scaling exponents
\cite{Ziff1986} are found to be $\tau = 187/91$ and $\tau' =
15/7$. In Fig.~\ref{Fig7} we present in logarithmic scales nodal
domain areas distribution $\langle n_s/n \rangle$ vs. $\langle
s/s_{min} \rangle$ obtained for the microwave rough billiard.  The
distribution $\langle n_s/n \rangle$ was constructed as normalized
to unity histogram with the bin equal to 1.  The areas $s$ of
nodal domains were calculated by summing up the areas of the
nearest neighboring grid sites having the same sign of the wave
function. In Fig. \ref{Fig7} the vertical axis $\langle n_s/n
\rangle = \frac{1}{N_T}\sum_{i=1}^{N_T}n_s^{(N)}/n^{(N)} $
represents the number of nodal domains $n_s^{(N)}$ of size $s$
divided by the total number of domains $n^{(N)}$ averaged over
$N_T=18$ wave functions measured in the range $250 \leq N \leq
435$.  In these calculations we used only the highest measured
wave functions in order to minimize the influence of boundary
domains on nodal domain areas distribution. Following Bogomolny
and Schmit \cite{Bogomolny2002}, the horizontal axis is expressed
in the units of the smallest possible area $s_{min}^{(N)}$,
$\langle s /s_{min} \rangle =
\frac{1}{N_T}\sum_{i=1}^{N_T}s/s_{min}^{(N)} $, where
$s_{min}^{(N)}=\pi(j_{01}/k_N)^2$ and $j_{01}\simeq 2.4048$ is the
first zero of the Bessel function $J_0(j_{01})=0$. The full line
in Fig. \ref{Fig7} shows the prediction of percolation theory
$\log_{10}(\langle n_s/n \rangle) = -\frac{187}{91}
\log_{10}(\langle s/s_{min} \rangle)$. In a broad range of
$\log_{10}(\langle s/s_{min} \rangle)$, approximately from 0.2 to
1.3, which is marked by the two vertical lines the experimental
results follow closely the theoretical prediction. Indeed, a least
squares fit $\log_{10}(\langle n_s/n \rangle) = a_2 -\tau
\log_{10}(\langle s/s_{min} \rangle)$ of the experimental results
lying within the vertical lines yields the scaling exponent $\tau
= 1.99 \pm 0.14$ and $a_2=-0.05 \pm 0.04$, which is in a good
agreement with the predicted $\tau= 187/91 \simeq 2.05$.  The
dashed line in Fig. \ref{Fig7} shows the results of the fit.  In
the vicinity of $\log_{10}(\langle s/s_{min} \rangle) \simeq 1 $
and $1.2$ small excesses of large areas are visible. A similar
situation, but for larger $ \log_{10} ( s/s_{min})
> 4$, can be also observed  in the nodal domain areas distribution
presented in  Fig. 5 in Ref. \cite{Bogomolny2002} for the random
wave model. The exact cause of this behavior is not known but  we
can possible link it with the limited number of wave functions
used for the preparation of the distribution.

\begin{figure}[!]
\begin{center}
\rotatebox{270} {\includegraphics[width=0.5\textwidth,
height=0.6\textheight, keepaspectratio]{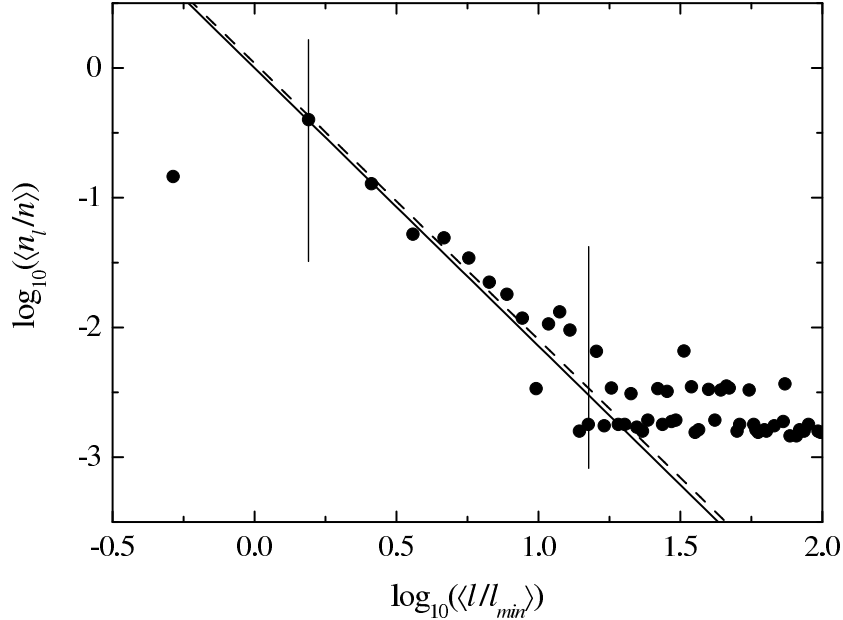}}
\caption{Distribution of nodal domain perimeters. Full line shows
the prediction of percolation theory $\log_{10}(\langle n_l/n
\rangle) =  -\frac{15}{7} \log_{10}(\langle l/l_{min} \rangle)$. A
least squares fit $\log_{10}(\langle n_l/n \rangle) = a_3 -\tau'
\log_{10}(\langle l/l_{min} \rangle)$ of the experimental results
lying within the range marked by the vertical lines yields $\tau'
= 2.13 \pm 0.23$ and $a_3=0.04 \pm 0.21$.  The result of the fit
is shown by the dashed line.} \label{Fig8}
\end{center}
\end{figure}

Nodal domain perimeters distribution $\langle n_l/n \rangle$ vs.
$\langle l/l_{min} \rangle$ is shown in logarithmic scales in Fig.
\ref{Fig8}. The distribution $\langle n_l/n \rangle$ was
constructed as normalized to unity histogram with the bin equal to
1 .   The perimeters of nodal domains $l$ were calculated by
identifying the continues paths of grid sites at the domains
boundaries. The averaged values $\langle n_l/n \rangle$ and
$\langle l/l_{min} \rangle$ are defined similarly as previously
defined $\langle n_s/n \rangle$ and $\langle s /s_{min} \rangle$,
e.g. $\langle l /l_{min} \rangle =
\frac{1}{N_T}\sum_{i=1}^{N_T}l/l_{min}^{(N)} $, where
$l_{min}^{(N)}=2\pi \sqrt{s_{min}^{(N)}/\pi} = 2\pi(j_{01}/k_N)$
is the perimeter of the circle with the smallest possible area
$s_{min}^{(N)}$. The full line in Fig. \ref{Fig8} shows the
prediction of percolation theory $\log_{10}(\langle n_l/n \rangle)
= -\frac{15}{7} \log_{10}(\langle l/l_{min} \rangle)$. Also in
this case the agreement between the experimental results and the
theory is good what is well seen in the range $0.2 <
\log_{10}(\langle l/l_{min} \rangle) < 1.2$, which is marked by
the two vertical lines. A least squares fit $\log_{10}(\langle
n_l/n \rangle) = a_3 -\tau' \log_{10}(\langle l/l_{min} \rangle)$
of the experimental results lying within the marked range  yields
$\tau' = 2.13 \pm 0.23$ and $a_3=0.04 \pm 0.21$.  The result of
the fit is shown in Fig. \ref{Fig8} by the dashed line. As we see
the scaling exponent $\tau'=2.13 \pm 0.23$ is close to the
exponent predicted by percolation theory $\tau'=15/7 \simeq 2.14$.
The above results clearly demonstrate that percolation theory is
very useful in description of the properties of wave functions of
chaotic billiards.

\begin{figure}[!]
\begin{center}
\rotatebox{270} {\includegraphics[width=0.5\textwidth,
height=0.6\textheight, keepaspectratio]{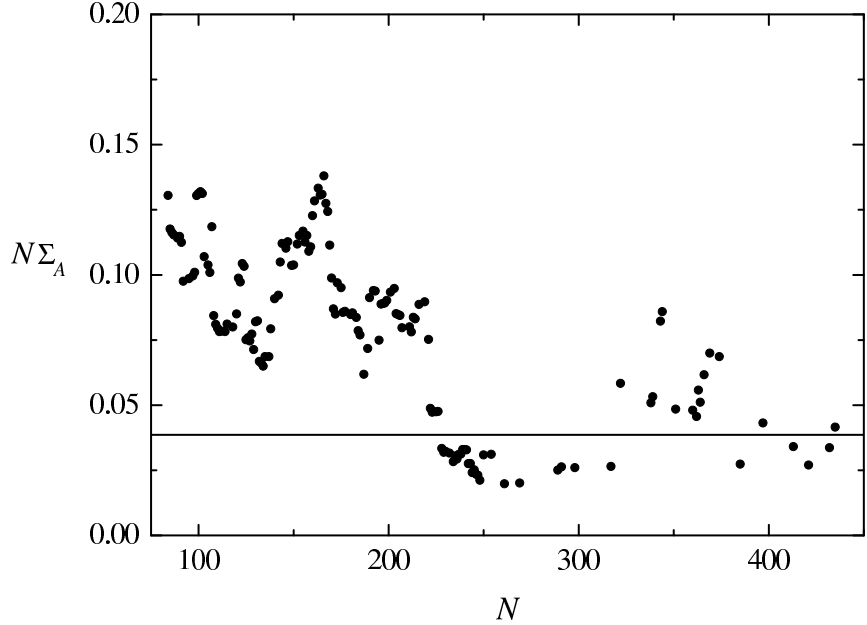}} \caption{The
normalized signed area distribution $N\Sigma_{A}$ for the chaotic
half-circular microwave rough billiard. Full line shows predicted
by the theory asymptotic limit $N\Sigma_{A} \simeq 0.0386 $, Blum
{\it et al.} \cite{Blum2002}.} \label{Fig9}
\end{center}
\end{figure}

Another important characteristic of the chaotic billiard is the
signed area distribution $\Sigma_{A}$  introduced by Blum {\it et
al.} \cite{Blum2002}. The signed area distribution is defined as a
variance: $\Sigma_{A}=\langle(A_{+} - A_{-})^2\rangle/A^2$, where
$A_{\pm}$ is the total area where the wave function is positive
(negative) and $A$ is the billiard area. It is predicted
\cite{Blum2002} that the signed area distribution should converge
in the asymptotic limit to $\Sigma_A \simeq 0.0386 N^{-1}$. In
Fig. \ref{Fig9} the normalized signed area distribution $
N\Sigma_{A}$ is shown for the microwave rough billiard. For lower
states $80\leq N \leq 250 $ the points in Fig. \ref{Fig9} were
obtained by averaging over 20 consecutive eigenstates while for
higher states $N
>250$ the averaging over 5 consecutive eigenstates was applied. For low
level numbers $N < 220$ the normalized distribution $N\Sigma_{A}$
is much above the predicted asymptotic limit,  however, for
$220<N\leq 435$ it more closely approaches the asymptotic limit.
This provides the evidence that the signed area distribution
$\Sigma_{A}$ can be used as a useful criterion of quantum chaos. A
slow convergence of $N\Sigma_{A}$ at low level numbers $N$ was
also observed for the Sinai and stadium billiards \cite{Blum2002}.
In the case of the Sinai billiard this phenomenon was attributed
to the presence of corners with sharp angles. According to Blum
{\it et al.} \cite{Blum2002} the effect of corners on the wave
functions is mainly accentuated at low energies. The half-circular
microwave rough billiard also possesses two sharp corners and they
can be responsible for a similar behavior.

In summary, we measured the wave functions of the chaotic rough
microwave billiard up to the level number $N=435$.  Following the
results of percolationlike model proposed by \cite{Bogomolny2002}
we confirmed that the distributions of the areas $s$ and the
perimeters $l$ of nodal domains have power behaviors $n_s \propto
s^{-\tau}$ and $n_l \propto l^{-\tau'}$, where scaling exponents
are equal to $\tau = 1.99 \pm 0.14$ and $\tau'=2.13 \pm 0.23$,
respectively. These results are in a good agreement with the
predictions of percolation theory \cite{Ziff1986}, which predicts
$\tau= 187/91 \simeq 2.05$ and $\tau'=15/7 \simeq 2.14$,
respectively. We also showed that in the limit $N \rightarrow
\infty $ a least squares fit of the experimental data yields the
asymptotic number of nodal domains $\aleph_N/N \simeq 0.058 \pm
0.006$ that is close to the theoretical prediction $\aleph_N/N
\simeq 0.062$ \cite{Bogomolny2002}. Finally, we found out that the
signed area distribution $\Sigma_{A}$ approaches for high level
number $N$  theoretically predicted asymptotic limit
$0.0386N^{-1}$ \cite{Blum2002}.

Acknowledgments.  This work was partially supported by KBN grant
No. 2 P03B 047 24. We would like to thank Szymon Bauch for
valuable discussions.

\end{document}